\documentstyle[12pt,a4wide]{article}
\newcommand{\slk}{/\kern-6pt k}
\newcommand{\sll}{/\kern-4pt l}
\newcommand{\slp}{p\kern-5pt/}
\newcommand{\slq}{q\kern-5.5pt/}
\newcommand{\slv}{v\kern-5pt\raise1pt\hbox{$\scriptstyle/$}\kern1pt}
\newcommand{\as}{\alpha_s}
\newcommand{\gs}{g_s}
\newcommand{\eps}{\varepsilon}

\begin{document}

\thispagestyle{empty}
\begin{flushright}
MZ-TH/97-05\\
hep-ph/9702374\\
February 1997\\
\end{flushright}
\vspace{0.5cm}
\begin{center}
{\Large\bf Two-loop Anomalous Dimensions}\\[.3cm]
{\Large\bf of Heavy Baryon Currents}\\[.3cm]
{\Large\bf in Heavy Quark Symmetry\footnote{Talk given at the Triangle 
  Graduate School in Particle Physics,\\\hbox{\qquad}Prague, Czech Republic, 
  September 2--10, 1996}}\\[1.3cm]
{\large Stefan Groote}\\[1cm]
Institut f\"ur Physik, Johannes-Gutenberg-Universit\"at,\\[.2cm]
Staudinger Weg 7, D-55099 Mainz, Germany\\
\vspace{1cm}
\end{center}

\begin{abstract}
This talk presents details of the calculation of two-loop corrections to 
heavy baryon currents containing one heavy quark in leading order of the 
$1/m_Q$-expansion, i.e.\ in the limit of Heavy Quark Symmetry (HQS). The 
calculations lead to the determination of the anomalous dimension of these 
currents. I also discuss problems with different $\gamma_5$-schemes and 
their solution by the finite renormalization procedure.
\end{abstract}

\newpage

\section{Introduction}
Heavy baryons such as $\Lambda_c$, $\Sigma_c$, $\Lambda_b$ or $\Sigma_b$ 
containing one heavy quark can be successfully described by the Heavy Quark 
Effective Theory (HQET), especially in leading order of the 
$1/m_Q$-expansion, i.e.\ in the limit of Heavy Quark Symmetry (HQS) (for a 
review on HQET see~\cite{Wise}, for details on HQS see~\cite{Neubert}). 
For each of the ground-state baryons there are two independent current 
components, e.g.\footnote{For a complete description 
see~\cite{GrooteKoernerYakovlev}}
\begin{equation}
J_{\Lambda1}=[q^{iT}C\tau\gamma_5q^j]Q^k\eps_{ijk}\quad\mbox{and}\quad
J_{\Lambda2}=[q^{iT}C\tau\gamma_5\gamma_0q^j]Q^k\eps_{ijk},
\end{equation}
while the general structure of heavy baryon currents has the form
\begin{equation}
J=[q^{iT}C\tau\Gamma q^j]\Gamma'Q^k\eps_{ijk},
\end{equation}
where the index $T$ means transposition, $C$ is the charge conjugation 
matrix, $\tau$ is a matrix in flavour space, and $i$, $j$ and $k$ are colour 
indices. The effective static field of the heavy quark is denoted by $Q$, 
and $\Gamma$ and $\Gamma'$ are the Dirac structures of the different current 
parts.

In Fig.~1 are shown some examples of diagrams up to two-loop order which 
we calculated. The one- and two-loop diagrams can be grouped in three 
different classes, namely the heavy diquark system (diagrams (b), (c), (e), 
(f) and (g)), the light diquark system (diagrams (d) and (h)) and the 
irreducible vertex contribution (diagrams (i) and (j)). The first two can 
be handled in a manner similar to the meson case, and actually it turns 
out that the replacement $C_B\rightarrow C_F$ leads from the diquark to the 
meson quark-antiquark case (and not vice-versa). So the mesonic case can be 
seen as a byproduct of our calculations.

\section{Algorithmic calculations}
The calculation of all these diagrams can be executed by nearly the same 
procedure. The procedure is strictly algorithmic, so that the calculations 
could be automatized and programmed in MATHEMATICA. In the following I want 
to emphasize the main steps of this calculation, and I want to do this for 
the first two-loop heavy diquark diagram which is displayed in Fig.~2 (the 
arrows on the gluon lines only indicate the momentum flow).

\subsection{Colour structure}
Before starting the main calculation, the colour structure can be treated. 
In general it contains the Gell-Mann matrices, the epsilon tensor for 
resons of colourless, and the colour couplings of the gluons, which 
consists of a $\delta_{ab}$ for a single gluon and a factor $if_{abc}$ for 
a three-gluon vertex. In our case we get
\begin{equation}
q^iq^{j''}(T^a)_{j''}^{j'}(T^b)_{j'}^jQ^{k''}(T^a)_{k''}^{k'}
  (T^b)_{k'}^k\eps_{ijk}
  =\left(\frac{N_C+1}{2N_C}\right)^2q^iq^jQ^k\eps_{ijk}
  =:C_B^2q^iq^jQ^k\eps_{ijk}.
\end{equation}
$N_C$ is the number of colours, and the results can be expressed by the 
constants $C_A=N_C$, $C_B=(N_C+1)/2N_C$, $C_F=(N_C^2-1)/2N_C$,
$T_F=1/2$ and $N_F$, which denotes the number of flavours.

\subsection{Other structures}
Without the colour structure we get
\begin{eqnarray}
\tilde b_1^{(hl)}&=&\int\int(-i\gs\gamma_\alpha)\frac{i}{\slk}
  (-i\gs\gamma_\beta)\frac{i}{\sll}\Gamma\left(\frac{-i}{k^2}\right)
  \left(\frac{-i}{(k-l)^2}\right)\ \cdot\\&&\qquad\qquad\qquad\qquad\cdot\
  \frac{i}{\omega+kv}(-i\gs v^\beta)\frac{i}{\omega+lv}(-i\gs v^\alpha)
  \frac{d^nk}{(2\pi)^n}\frac{d^nl}{(2\pi)^n}\ =\nonumber\\
  &=&-\frac{\gs^4}{\omega^2}\int\int\slv\slk\slv\sll\Gamma
  \left(\frac{-1}{k^2}\right)^2\left(\frac{-1}{l^2}\right)
  \left(\frac{-1}{(k-l)^2}\right)\frac\omega{(\omega+kv)}
  \frac\omega{(\omega+lv)}\frac{d^nk}{(2\pi)^n}\frac{d^nl}{(2\pi)^n}.\nonumber
\end{eqnarray}
The four-momentum of the heavy quark is given by $q=m_Qv+p$, and we have
$\omega=pv$. This vertex correction can be split up into a {\em Dirac 
structure $\slv\gamma_\mu\slv\gamma_\nu\Gamma$}, a {\em momentum structure 
$k^\mu l^\nu$} and a {\em scalar integral} (also called {\em``race 
integral''})
\begin{eqnarray}
\hat b_1^{(hl)}&=&-\frac{\gs^4}{\omega^2}\int\int\left(\frac{-1}{k^2}\right)^2
  \left(\frac{-1}{l^2}\right)\left(\frac{-1}{(k-l)^2}\right)
  \frac\omega{(\omega+kv)}\frac\omega{(\omega+lv)}\frac{d^nk}{(2\pi)^n}
  \frac{d^nl}{(2\pi)^n}\ =\nonumber\\
  &=&\frac{\gs^4}{(4\pi)^n\omega^2}(-2\omega)^{2(n-4)}I_5(2,1,1,1,1).
\end{eqnarray}

\subsection{One- and two-loop functions}
With the help of the recursion formulas developed in~\cite{ChetyrkinTkachov,
SurguladzeTkachov,BroadhurstGrozin} it is possible to reduce each two-loop 
function $I_5$ of the HQET and each two-loop function $G_5$ of the massless 
QCD to corresponding one-loop functions $I_2$ and $G_2$, which dissolve 
into Euler's Gamma functions. The recursion formulas are one of the main 
parts of the developed program package, and we would like to thank David 
Broadhurst for providing us with a REDUCE implementation of these formulas.

\subsection{Momentum structure, covariant evaluation and operators}
By extracting the Dirac structure, the remaining integral receives 
corresponding Lorentz indices. It can be evaluated covariantly, in this 
case into
\begin{equation}
Ag^{\mu\nu}+Bv^\mu v^\nu.
\end{equation}
The parts $A$ and $B$ can be obtained by contracting the integral with the 
corresponding dual tensors. This contraction pairs the momentum structure 
to the scalar products
\begin{equation}
kl=\frac12(k^2+l^2-(k-l)^2)\quad\mbox{and}\quad(kv)(lv),\quad\mbox{resp.}
\end{equation}
The scalar products ``operate'' on the entries of the two-loop function
(e.g., $k^2$ decreases the first entry by one unit), and can therefore be 
identified with operators on the different entries of the two-loop 
functions.

\subsection{Dirac structure and its reduction}
The covariant evaluation in combination with the Dirac structure gives
\begin{equation}
\tilde b_1^{(hl)}=\slv\gamma_\mu\slv\gamma_\nu\Gamma(Ag^{\mu\nu}+Bv^\mu v^\nu)
  =A\slv\gamma_\mu\slv\gamma^\mu\Gamma+B\slv\slv\slv\slv\Gamma
  =((2-n)A+B)\Gamma.
\end{equation}
For the heavy diquark system the matrix $\Gamma$ will always remain on the 
edge of the structure. In contrast to this, for the light diquark system 
the Dirac structure can be reduced to one of the main structures
\begin{eqnarray}
\Gamma_0&:=&\Gamma,\qquad\Gamma_1\ :=\ \gamma_\mu\slv\Gamma\slv\gamma^\mu,
  \qquad\Gamma_2\ :=\ \gamma_\mu\gamma_\nu\Gamma\gamma^\nu\gamma^\mu,\\
\Gamma_3&:=&\gamma_\mu\gamma_\nu\gamma_\rho\slv\Gamma\slv\gamma^\rho
  \gamma^\nu\gamma^\mu\quad\mbox{and}\quad\Gamma_4\ :=\ \gamma_\mu\gamma_\nu
  \gamma_\rho\gamma_\sigma\Gamma\gamma^\sigma\gamma^\rho\gamma^\nu\gamma^\mu.
  \nonumber
\end{eqnarray}

\section{Result of the perturbation series}
The result of the perturbation series to second order for the different 
ground-state hadron currents can be written in quite a compact form, if we 
make some slight assumption for $\Gamma$. In this case, we can split off a 
scalar valued {\em vertex function $V$}.

\subsection{Introducing $n_\gamma$ and $s_\gamma$}
If $\Gamma$ is totally antisymmetrized,
$\Gamma\in\{1,\gamma^\mu,\gamma^{[\mu}\gamma^{\nu]},\ldots\}$, we get
\begin{equation}
\gamma_\mu\Gamma\gamma^\mu=(-1)^{n_\gamma}(n-2n_\gamma)\Gamma\quad
  \mbox{and}\quad\gamma_0\Gamma\gamma_0=(-1)^{n_\gamma}s_\gamma\Gamma,
\end{equation}
where $n$ is the space-time dimension, $n_\gamma$ the number of Dirac 
matrices and $s_\gamma$ denotes the number of $\gamma_0$'s in $\Gamma$, 
i.e.\ being $+1$ for an even and $-1$ for an odd number of $\gamma_0$'s.
Problems will emerge from the appearence of $\gamma_5$, but I will 
postpone the discussion of this to the end of the talk.

\subsection{Vertex function for the heavy baryon currents}
Without looking at the parts from different classes of diagrams, the vertex 
function for the heavy baryon current in its generalized form is given by
\begin{equation}
V=1+2\Delta V^{(hl)}+\Delta V^{(ll)}+\Delta V^{(ir)}
\end{equation}
This expression can be evaluated in powers of $1/\eps$ with $\eps=(4-n)/2$ 
and shows singularities which can be canceled by means of the minimal 
subtraction scheme (MS). Indeed, the evaluation in
\begin{equation}
\frac1{\eps'}:=\frac1\eps-\gamma_E+\ln(4\pi)=\frac1\eps\ln(4\pi e^{\gamma_E})
\end{equation}
is rather more simple. This evaluation allows for the application of the 
modified minimal subtraction scheme ($\overline{\rm MS}$). By using a 
general gauge parameter $a$ ($a=1$ means Feynman gauge) for the one-loop 
contributions, we get
\begin{equation}
V=\sum_{m=0}^\infty\left(\frac{\as}{4\pi}\right)^mV_m
  =\sum_{m=0}^\infty\sum_{k=0}^m\left(\frac{\as}{4\pi}\right)^m
  \frac1{\eps'^k}V_m^k\quad\mbox{with}
\end{equation}
\begin{eqnarray}
V_1^1&=&C_B\Big((n_\gamma-2)^2+3a-1\Big),\nonumber\\
V_1^0&=&C_B\Big(2n_\gamma^2-6n_\gamma+3+a+(n_\gamma-2)s_\gamma\Big),\nonumber\\
V_2^2&=&\frac{C_B}6\Big(3C_B((n_\gamma-2)^2+2)^2
  -C_A(11n_\gamma^2-44n_\gamma+51)\,+\nonumber\\&&
  +4N_FT_F(n_\gamma-3)(n_\gamma-1)\Big)\quad\mbox{und}\\
V_2^1&=&\frac{C_B}{36}\Big(9C_B((n_\gamma-2)(13n_\gamma^3-70n_\gamma^2
  +126n_\gamma-60+4(n_\gamma^2-4n_\gamma+7)s_\gamma)+32\zeta(2))
  \,+\nonumber\\&&
  -C_A(18n_\gamma^4-144n_\gamma^3+289n_\gamma^2+128n_\gamma-516+72\zeta(2))
  \,+\nonumber\\&&
  -72C_F(n_\gamma-3)(n_\gamma-1)-4N_FT_F(n_\gamma^2-16n_\gamma+24)
  \Big).\nonumber
\end{eqnarray}
$V_2^0$ is omitted, because it has no significance for the following 
calculations.

\section{The renormalization}
The vertex function still contains UV-singularities. It is a bare quantity,
from now on denoted by $V^0$, which has to be renormalized. This 
renormalization is done with the help of a renormalization factor. 
Dimensional regularization is used from the beginning, what remains is the 
subtraction with the aid of the modified minimal subtraction scheme, as 
indicated above. In the following we simplify the notation by using the 
term $\eps$ instead of $\eps'$, but we keep in mind that we have used the 
$\overline{\rm MS}$-scheme.

\subsection{From the perturbation series to the renormalization factor}
The divergencies of $V^0$ will be absorbed by the renormalization factor 
$Z_V$. We have to state therefore that $Z_V^{-1}V^0=V$ is a finite 
quantity. By expanding the renormalization factor $Z_V$ in a double series
\begin{equation}
Z_V=1+\sum_{m=1}^\infty\sum_{k=1}^m\left(\frac{\as}{4\pi}\right)^m
  \frac1{\eps^k}Z_m^k=\sum_{k=0}^\infty\frac1{\eps^k}Z^k
  \quad\mbox{with}\quad Z^0=1,
\end{equation}
a comparison of the coefficients up to second order gives
\begin{equation}
Z_1^1=V_1^1,\quad Z_2^2=V_2^2\quad\mbox{and}\quad Z_2^1=V_2^1-V_1^0V_1^1.
\end{equation}

\subsection{Renormalization of coupling and gauge parameter}
Up to now we have left out the renormalization of the coupling 
constant~$\as$ and the gauge parameter~$a$,
\begin{equation}
\as^0=\as\left(1+\frac{\as}{4\pi\eps}\delta_\alpha+O(\as^2)\right),\qquad
a^0=a\left(1+\frac{\as}{4\pi\eps}\delta_a+O(\as^2)\right).
\end{equation}
Looking at terms up to second order in perturbation theory, we only need 
to correct the terms of first order. In this way we get corrections for 
the coefficient $Z_2^1$ of the renormalization factor $Z_V$, namely
\begin{equation}
Z_2^1=V_2^1-V_1^0V_1^1+\delta_\alpha V_1^0+\delta_a\tilde V_1^0,
\end{equation}
where $\tilde V_1^0$ is the gauge dependent part of $V_1^0$.

\subsection{From the renormalization factor to the anomalous dimension}
The renormalized coupling is an object in $n$ space-time dimensions. 
Therefore, it gets a double dependence in the renormalization scale~$\mu$,
\begin{equation}
\as=\as(\mu,\eps)=\as(\mu)\mu^{2\eps}\quad\mbox{($\as(\mu)$ is the 
  coupling in $4$ space-time dimensions).}
\end{equation}
The beta-function for the coupling and the gauge, two essential tools for the 
renormalization group equation, can then be calculated as
\begin{equation}
\beta=\frac{d\ln\as(\mu,\eps)}{d\ln\mu}
  =\frac{d\ln\as(\mu)}{d\ln\mu}+2\eps,\qquad
\beta_a=\frac{d\ln a(\mu)}{d\ln\mu}.
\end{equation}
This can be used for the calculation of the anomalous dimension $\gamma$,
\begin{eqnarray}
\gamma Z&=&\frac{dZ(\as(\mu),a(\mu);\eps)}{d\ln\mu}
  \ =\ \frac{\partial Z}{\partial\ln\as}\frac{d\ln\as(\mu)}{d\ln\mu}
  +\frac{\partial Z}{\partial\ln a}\frac{d\ln a(\mu)}{d\ln\mu}\ =\nonumber\\
  &=&\frac{\partial Z}{\partial\ln\as}(\beta-2\eps)
  +\frac{\partial Z}{\partial\ln a}\beta_a.
\end{eqnarray}
By evaluating the finite quantities $\beta$, $\beta_a$ and $\gamma$ in a 
perturbation series, starting at first order, the comparison of the 
coefficients leads to a determining equation for the coefficients of the 
anomalous dimension,
\begin{equation}
\gamma_m=-2mZ_m^1,
\end{equation}
and on the other hand to a class of consistency checks, which can be 
revolved to the coefficients of the perturbation series, e.g.
\begin{equation}
2V_2^2=V_1^1V_1^1+\delta_\alpha V_1^1+\delta_a\tilde V_1^1
\end{equation}
where $\tilde V_1^1$ again denotes the gauge dependent part of $V_1^1$. For 
the derivation we should note that
\begin{equation}
\delta_\alpha=\frac12\beta_1=-\frac{11}3C_A+\frac43N_FT_F\quad\mbox{and}
  \quad\delta_a=\frac12(\beta_a)_1=\frac{13-3a}6C_A-\frac43N_FT_F.
\end{equation}

\section{The anomalous dimension of the\hfil\break
  generalized heavy baryon current}
To renormalize the current, we also have to include the renormalization of 
the outer legs, because we have
\begin{equation}
J^0=[q^{0T}C\tau V^0\Gamma q^0]\Gamma'Q^0=Z_qZ_Q^{1/2}Z_VJ=Z_JJ\quad
  \Rightarrow\quad\gamma_J=2\gamma_q+\gamma_Q+\gamma_V.
\end{equation}
By taking this into account~\cite{BroadhurstGrozin,JonesEtc}, to first order 
we get
\begin{equation}
\gamma_J=\frac{\as}{4\pi}\Big(-2C_B((n_\gamma-2)^2+3a-1)+3C_F(a-1)\Big)
  +O(\as^2).
\end{equation}
For the values $C_B=2/3$ and $C_F=4/3$ of $SU(3)_C$, this expression is 
gauge independent. So we can use these values, $C_A=3$ and $T_F=1/2$ and 
restrict ourselves to the Feynman gauge.

\subsection{Different $\gamma_5$-schemes}
The problem in presenting the final result consists in the occurence of 
different schemes for $\gamma_5$.
\bigskip\\
\parbox{9truecm}{In the scheme due to 't~Hooft, 
Veltman, Breitenlohner and Maison~\cite{tHooftVeltman,BreitenlohnerMaison},
an additional $\gamma_5$ in the vertex chances $n_\gamma$ to $4-n_\gamma$
and $s_\gamma$ to $-s_\gamma$. For the naively anticommuting 
$\gamma_5$-scheme~\cite{Kreimer}, both defining equations for $n_\gamma$ 
and $s_\gamma$ change their general sign, but this will cancel for all main 
structures $\Gamma_i$. The values for the parameters for different heavy 
baryons are displayed in the enclosed table.}
\hfill
\parbox{6truecm}{\begin{tabular}{|l|c|c|l|}\hline
$\Gamma$&$n_\gamma$&$s_\gamma$&baryon\\\hline\hline
$\gamma^{\rm AC}_5$&$0$&$+1$&$\Lambda_1$\\\hline
$\gamma^{\rm AC}_5\gamma_0$&$1$&$-1$&$\Lambda_2$\\\hline
$\vec\gamma$&$1$&$+1$&$\Sigma_1,\Sigma^*_1$\\\hline
$\gamma_0\vec\gamma$&$2$&$-1$&$\Sigma_2,\Sigma^*_2$\\\hline\hline
$\gamma^{\rm HV}_5$&$4$&$-1$&$\Lambda_1$\\\hline
$\gamma^{\rm HV}_5\gamma_0$&$3$&$+1$&$\Lambda_2$\\\hline
\end{tabular}}

\subsection{Values for the anomalous dimension}
For the anomalous dimension of different baryon currents and different 
schemes we get
\begin{eqnarray}
\gamma_{\Lambda1}^{\rm AC}
  &=&-8\left(\frac{\alpha_s}{4\pi}\right)+\frac19(16\zeta(2)+40N_F-796)
  \left(\frac{\alpha_s}{4\pi}\right)^2,\nonumber\\ 
\gamma_{\Lambda2}^{\rm AC}
  &=&-4\left(\frac{\alpha_s}{4\pi}\right)+\frac19(16\zeta(2)+20N_F-322)
  \left(\frac{\alpha_s}{4\pi}\right)^2,\nonumber\\
\gamma_{\Sigma1}
  &=&-4\left(\frac{\alpha_s}{4\pi}\right)+\frac19(16\zeta(2)+20N_F-290)
  \left(\frac{\alpha_s}{4\pi}\right)^2,\\
\gamma_{\Sigma2}
  &=&-\frac83\left(\frac{\alpha_s}{4\pi}\right)+\frac1{27}(48\zeta(2)+8N_F+324)
  \left(\frac{\alpha_s}{4\pi}\right)^2,\nonumber\\
\gamma_{\Lambda1}^{\rm HV}
  &=&-8\left(\frac{\alpha_s}{4\pi}\right)+\frac19(16\zeta(2)-24N_F+260)
  \left(\frac{\alpha_s}{4\pi}\right)^2,\nonumber\\
\gamma_{\Lambda2}^{\rm HV}
  &=&-4\left(\frac{\alpha_s}{4\pi}\right)+\frac19(16\zeta(2)-12N_F+206)
  \left(\frac{\alpha_s}{4\pi}\right)^2.\nonumber
\end{eqnarray}
The ``naive non-Abelianization'' recipe, the estimate of the total result 
by calculating the quark-loop diagram and replacing $N_F$ by $N_F-33/2$, 
does work for most of the cases with an accuracy of 15 to 30\%. Only in the 
case of the current $J_{\Sigma2}$ it seems to fail. Note, however, that in 
this case the anomalous dimension is one order of magnitude smaller than in 
all other cases.

\subsection{The finite renormalization}
The difference of the anomalous dimension for the two different 
$\gamma_5$-schemes turns out to be a correction of second order which is 
proportional to the coefficient $\beta_1$ of the beta-function of the 
coupling. So we might think about connecting the already renormalized 
currents by some finite renormalization factor~\cite{TruemanLarin}, namely 
\begin{equation}
J^{\rm AC}(\mu)=Z_\Gamma J^{\rm HV}(\mu).
\end{equation}
So we get
\begin{equation}
\gamma_\Gamma=\frac{d\ln Z_\Gamma}{d\ln\mu}
  =\gamma_J^{\rm HV}-\gamma_J^{\rm AC}
  =:O_\Gamma\left(\frac{\as}{4\pi}\right)^2.
\end{equation}
Indeed, the ansatz $Z_\Gamma=A_\Gamma(\as(\mu)/4\pi)$ leads to 
$A_\Gamma\beta_1=O_\Gamma$ with
\begin{eqnarray}
O_{\gamma_5}&=&\frac{16}3C_B(11C_A-4N_FT_F),\nonumber\\
O_{\gamma_5\gamma_i}\ =\ O_{\gamma_5\gamma_0}
  &=&\frac83C_B(11C_A-4N_FT_F)\quad\mbox{and}\\[7pt]
O_{\gamma_5\gamma_i\gamma_j}\ =\ O_{\gamma_5\gamma_i\gamma_0}&=&0,\nonumber
\end{eqnarray}
and therefore
\begin{equation}
Z_{\gamma_5}=1-8\frac{\as}{4\pi}C_B,\quad
Z_{\gamma_5\gamma_i}=Z_{\gamma_5\gamma_0}=1-4\frac{\as}{4\pi}C_B\quad
\mbox{and}\quad Z_{\gamma_5\gamma_i\gamma_j}=Z_{\gamma_5\gamma_i\gamma_0}=0.
\end{equation}
The vertex structures $\Gamma=\gamma_5\gamma_i$, $\gamma_5\gamma_i\gamma_j$ 
and $\gamma_5\gamma_i\gamma_0$ are included for the sake of completeness, 
although they are not needed for the baryonic $s$-wave states discussed in 
this talk. The meson case, already treated by Trueman and 
Larin~\cite{TruemanLarin}, can again be obtained by the replacement 
$C_B\rightarrow C_F$.\\[1truecm]
\noindent{\large \bf Acknowledgments:}\smallskip\\
This work was partially supported by the BMBF, FRG, under contract 06MZ566.  
Moreover, I am grateful to the DFG, FRG, for financial support. I would 
like to thank J.G.~K\"orner and O.I.~Yakovlev for joining me in doing the 
presented calculations.

\vspace{1truecm}
\centerline{\Large\bf Figure Captions}
\vspace{.5cm}
\newcounter{fig}
\begin{list}{\bf\rm Fig.\ \arabic{fig}:}{\usecounter{fig}
\labelwidth1.6cm\leftmargin2.5cm\labelsep.4cm\itemsep0ex plus.2ex}
\item Examples for corrections up to two-loop order to the baryonic vertex.
Double lines indicate heavy quarks, single lines indicate massless quarks, 
the springs stand for gluons
\item The first heavy diquark diagram (Fig.~1(e)) with momentum labels
\end{list}

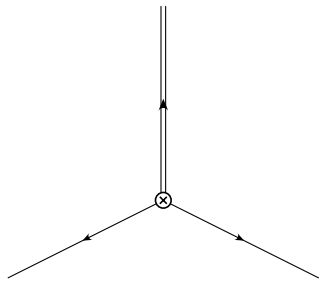
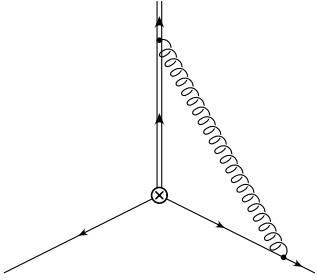
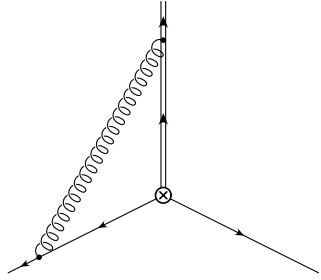
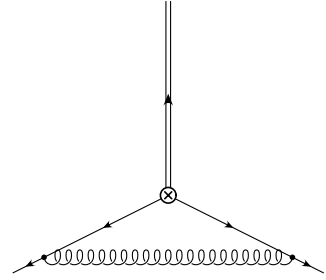
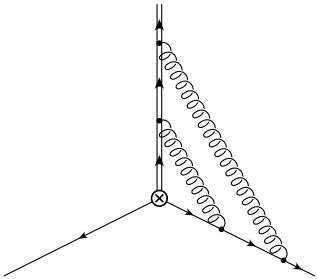
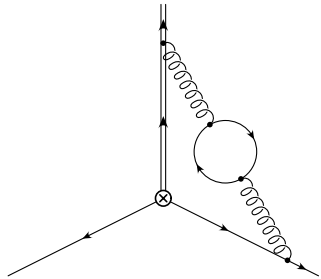
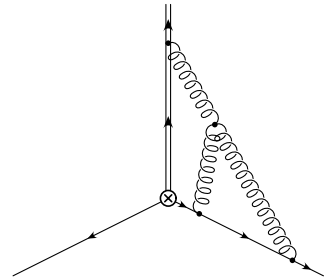
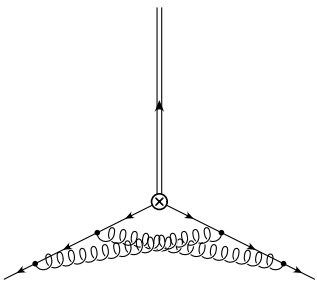
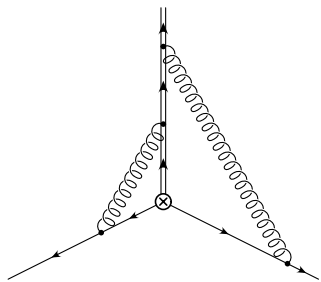
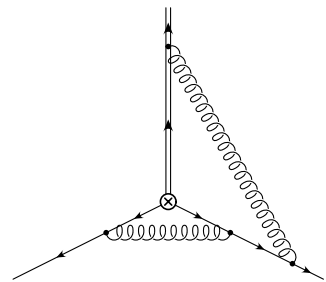
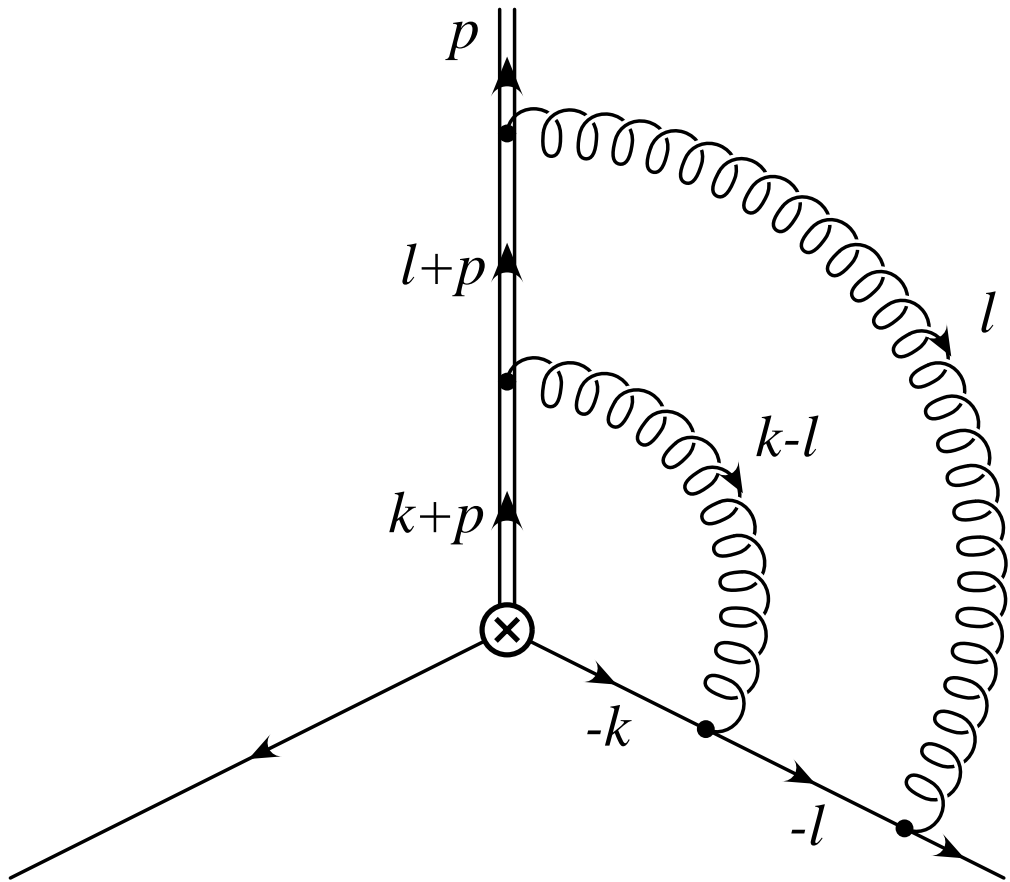
\end{document}